# A Topological Pattern of Urban Street Networks: Universality and Peculiarity

Bin Jiang

Department of Land Surveying and Geo-informatics
The Hong Kong Polytechnic University, Hung Hom, Kowloon, Hong Kong
Email: bin.jiang@polyu.edu.hk

**Abstract:** In this paper, we derive a topological pattern of urban street networks using a large sample (the largest so far to the best of our knowledge) of 40 U.S. cities and a few more from elsewhere of different sizes. It is found that all the topologies of urban street networks based on street-street intersection demonstrate a small world structure, and a scale-free property for both street length and connectivity degree. More specifically, for any street network, about 80% of its streets have length or degrees less than its average value, while 20% of streets have length or degrees greater than the average. Out of the 20%, there are less than 1% of streets which can form a backbone of the street network. Based on the finding, we conjecture that the 20% streets account for 80% of traffic flow, and the 1% streets constitute a cognitive map of the urban street network. We illustrate further a peculiarity about the scale-free property.

**Keywords:** urban street networks, small world, scale free, self-organizing cities, and cognitive maps

## 1. Introduction

In the last decade, topological analysis has been widely adopted to uncover patterns or structures from various real world systems including information, social, technology and biological. This new wave of interest revival is mainly triggered by the discovery of small world and scale free networks [1, 2] and the increasing availability of the various real world network datasets like the Internet and the World Wide Web. The emergence of geographic information systems (GIS) has generated an increasing amount of geospatial data of urban street networks. An urban street network is often perceived as a graph whose edges represent street segments and vertices represent segment intersections (or junctions). This network is the traditional representation for transportation modeling [3], but it is not suitable for uncovering structures or patterns. On the surface such a network has a pretty simple structure. To uncover structures or patterns from the network representation is like trying to perceive an image in terms of pixels rather than things or objects, i.e., the street segments are to a network what the pixels are to an image. To perceive a street network, we need to merge individual segments into meaningful streets, just as we merge individual pixels into meaningful objects.

Two different approaches for merging individual street segments into meaningful streets have been adopted in the literature. The first is based on the same street name to merge individual street segments, and this is the so called named streets approach [4]. The second relies on a good continuation to merge adjacent street segments together to form meaningful streets. The streets are also called strokes [5] based on perceptual grouping, a fundamental concept of Gestalt psychology [6], which has wide applications in human and computer vision for pattern recognition [7]. Based on the streets, a graph can be formed by taking the streets as vertices and street intersections as edges of the graph for topological analysis. However, the two approaches have their pros and cons. The first approach must rely on very accurate information about street names for the merging process, a semantic approach in essence. Unfortunately, even the most comprehensive databases cannot guarantee that every street segment has a valid name, not to say that incorrect names are assigned to some street segments. This incomplete or inaccurate nature of the database is a big impediment to adopt the first approach. In this respect, the second approach shows a complementary advantage, since valid names are not necessary for the merging process.

In this paper, we adopt the second approach to form streets and then check street-street intersections to form a topology of an urban street network for analysis. We apply the topological analysis to a very large sample of U.S. cities as well as a few more from elsewhere in order to illustrate both universality and peculiarity of the topological pattern of urban street networks. In this connection, primary studies have been done by some researchers. For instance, based on the named street approach, Jiang [8] re-examined scale-free property of degree distribution, and confirmed a power law distribution that revised their initial conclusion [4]. Both [9] and [10] have studied some city samples, and pointed out a general tendency of power law distributions, but no particular form which the power law distribution might take. Criticisms have been made of the studied samples in drawing a reliable conclusion as one-square mile areas have been used [11]. The street networks from the studied samples seem incomplete. For example, the city of Los Angeles contains only 240 streets, but it involves 57752 streets according to our sample (see Table 1 late in this paper). Carvalho and Penn [12] have also illustrated the power law distributions of open space linear segments derived from a very big city sample. The similar scaling property has also been found in nationwide road networks [13]. This paper is intended to be an in-depth study on the street-street intersection topology based on the largest city sample extracted from the Topologically Integrated Geographic Encoding and Referencing (TIGER) database developed by the U.S. Census Bureau. We will illustrate an emergent pattern in terms of street-street intersection topology. That is, for any street network, about 80% of streets have a length or degrees less than the average value of its network, while 20% of streets have length or degrees greater than the average. Out of the 20%, there are less than 1% of streets which can form a backbone of the street network. We will demonstrate how this illustrated topological pattern is universal for different cities, even for different parts of a city (of a reasonable size). Apart from the universality, we will also illustrate some peculiarity on the scaling property, e.g., some street-street intersection topologies have two (or even more) exponents.

The remainder of this paper is structured as follows. Section two briefly introduces the TIGER database, a reliable database for topological analysis, and describes in detail how the datasets of urban street networks are pre-processed for extracting street-street intersection topologies. In section three, we compute and analyze topological measures for the largest city sample, and illustrate the universal pattern about topologies of urban street networks. Finally in section four, we point out that the findings have far reaching implications towards the understanding of cognitive maps and self-organizing cities.

**2. Main data source and data processing based on perceptual grouping**
The main data source used in the study is the TIGER database, developed by the U.S. Census Bureau in the 1970s. It is mainly based on the theories of topology and graph theory for encoding the location of geographic objects such as streets, rivers, railroads and other features and their relationships. The topological structure of the TIGER ensures no duplication of these geographic features or areas, thus a very economic data organization. Urban street networks are stored in the format of Digital Line Graphs, which are topologically structured. The network database is well maintained and updated, so it is a reliable database for this kind of topological analysis. Taking Los Angeles for example, it contains 93 isolated street segments out of 225346 total street segments, less than 0.04%. On the other hand, it contains 7699 segments (over 3%) that have no appropriate street names. These segments with missing names appear to be with road junctions, thus deleting them would significantly distort the topological patterns. For this reason, we believe that the stroke-based approach tends to be a more robust model for the analysis.

We selected 40 cities (or urban areas to be more general) ranked between the first and 160th according to population. The 40 cities were deliberately chosen from the database for the study. We first chose the top 10 cities, and second, 10 cities between 91st and 100th, and then another ten in the middle of the top 100, ranging from 56th to 65th. After extensive study and exploration, another 10 smaller cities were chosen from a range of 151-160. Therefore the city sample represents a diverse set in terms of both size and their geographic locations. The dataset can be downloaded from the U.S. Census Bureau website [14]. For convenience, we downloaded it from the ESRI ArcData site [15], which has a more friendly interface. The main advantage of using the data is that we can take an entire urban street network rather than a partial one for analysis. To ensure street networks that are truly in an urban area, we used an urban area boundary data layer of 2000 to crop the street networks. The two layers (i.e., line features – roads, and urban areas of 2000) are overlapped to crop the street networks within individual urban areas. This guarantees that the datasets are truly urban street networks. The cropped urban street networks have naturally evolved boundaries, and some of them have very interesting animal-like shapes (see Figure 1 for an example). To this point, the reader may have understood that the street networks are not necessary to be within cities but within urban areas for sure.

The cropped urban street networks were then pre-processed in order to form individual streets based on the Gestalt principle of good continuation. Good continuation refers to an innate tendency to perceive a line continuing in its established direction. By checking the deflection angle from one segment to another, it will help determine whether or not two adjacent line segments are perceived as a street. A perfect continuation implies that the deflection angle between one street segment and its adjacent one is zero. However, the condition can be relaxed dramatically. We chose 70 degrees as a threshold for determining continuity in the study. This is based on our observation that this degree threshold is likely to detect ring roads or semi-ring roads which often appear in cities. The grouping process takes an ordinary street segments-based shape file as input, and generates a shape file that is streets-based. The process can be described as follows: for every street segment, we trace its connected segments, and concatenate the segment and an adjacent one with the smallest deflection angle. Note: this process should be done at both ends of every street segment.

Next we want to do a cleaning process to ensure that all streets are interconnected, or isolated ones are taken away. For this purpose, we select one street as a root, and adopt the well-known Breadth-First Search algorithm [16] to explore all those streets that directly or indirectly connect to the root street. Clearly those that are not connected either directly or indirectly to the root street are isolated ones, which should be deleted. It should be noted that the percentage of isolated streets tends to be a very small portion as mentioned above.

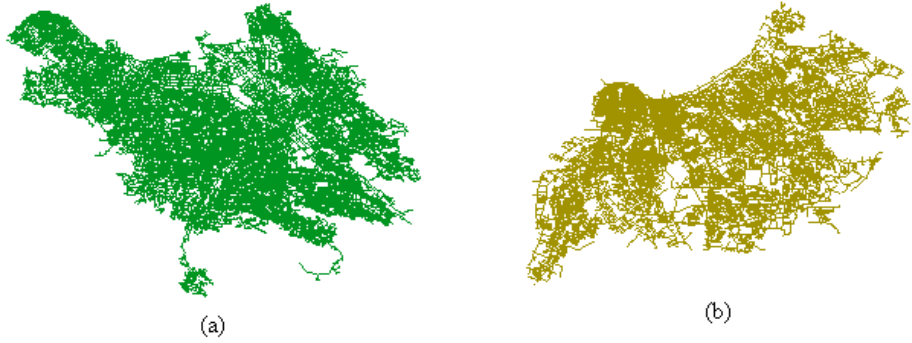

Figure 1: Urban street networks of Sunnyvale (a) and Louisville (b) from the city sample

## 3. Topological Analysis

After the above grouping and cleaning pre-process, all isolated street(s) are deleted for the following process. We compare every street to every other street, to see if they are intersected, in order to form a topology, or connected graph. In general, a graph (*G*) consists of a finite set of vertices (or nodes) $V = \{v_1, v_2, ... v_n\}$ (where the number of nodes is *n*) and a finite set of edges (or links) *E*, which is a subset of the Cartesian product $V \times V$. The connected graph can be represented as a matrix *R(G)*, whose element $r_{ij}$ is 1 if intersected, and 0 otherwise. Formally it is represented as follows:

$$R_{ij} = \begin{cases} 1 & \text{if } i \text{ and } j \text{ are intersected} \\ 0 & \text{otherwise} \end{cases} \quad (1)$$

It should be noted that this matrix *R(G)* is symmetric, i.e. $\forall r_{ij} \Rightarrow r_{ij} = r_{ji}$, and that all diagonal elements of *R(G)* are equal to zero. From a computational point of view, we compare each street to only those streets within a same envelop, a rectangular area that covers a street.

Before illustrating in detail the result of topological analysis, let's first briefly introduce three topological measures, namely degree, path length and clustering coefficient. For a vertex, its degree is the number of other vertices directly connected to it. The average of degree of all individual nodes is that of the graph. Formally it is defined by:

$$m(G) = \frac{1}{n} \sum_{j=1}^{n} R_{ij} \quad (2)$$

Path length of a graph is to measure how compact a graph is. For a graph, if every vertex is connected to every other, then it is very compact, thus the smallest path length. On the other hand, if all the vertices are connected as a linear chain, then it is organized in a very loose way with the largest path length. Formally path length is defined as:

$$\ell(G) = \frac{1}{n} \sum_{i=1}^{n} \sum_{j=1}^{n} d(i, j), \quad (3)$$

where $d(i, j)$ denotes the distance between two vertices *i* and *j*, which is the minimum length of the paths that connect the two vertices, i.e., the length of a *graph geodesic*.

Clustering coefficient is to measure the clustering degree of a graph. It can be seen from how clustered each individual vertex is, which can be defined as the probability that two neighbours of a given node are linked together, i.e., a ratio of the number of actual edges to that of possible edges. The average clustering coefficient of individual vertices is that of their graph, i.e.,

$$c(G) = \frac{1}{n} \sum_{i=1}^{n} \frac{\# \ of \ actual \ edges}{\# \ of \ possible \ edges}, \qquad (4)$$

For topological analysis, we first compute the measures for the topologies of the 40 urban street networks, and the results are presented in Table 1. We found that all the topologies without a single exception show a small world structure, with path length pretty close to that of their random counterparts. This result significantly deviates from what has been illustrated in the previous studies [9, 10], where path length tends to be slightly greater than their random counterparts. The reason, we suspect, is due to the fact that that the two studies used one-square mile area or part of a city in their city samples, and their street networks are derived from scanned binary images [17]. We further examined the distributions of degree and found they all exhibit power law distributions, i.e., $p(x) \sim cx^{-\alpha}$. Figure 2 demonstrates log-log plots, whose x-axis and y-axis represent the logarithms of degree and cumulative probability. We can remark that most of the log-log curves are pretty close to a straight line with an exponent around 2.0, thus a clear indication of power laws. The exponent for all the log-log curves is computed according to the formula suggested by [18].

Table 1: The 40 U.S. urban street networks and their topological measures
(Rank = ranking according to population, α = power law curve exponent, partition = percentage around the average degree, n = number of streets, m = average degree of streets, $\ell$ = path length, $\ell \ rand$ = path length of the random counterpart)

| Rank | City | State | α | Partition | n | m | $\ell$ | $\ell \ rand$ |
|---|---|---|---|---|---|---|---|---|
| 2 | Los Angeles | California | 2.0 | 4(79%) | 57752 | 4.2 | 7.3 | 7.6 |
| 9 | Phoenix | Arizona | 2.1 | 4(85%) | 50683 | 3.4 | 7.7 | 9.0 |
| 3 | Chicago | Illinois | 2.0 | 5(82%) | 44421 | 4.6 | 6.6 | 7.0 |
| 4 | Houston | Texas | 2.1 | 4(81%) | 42873 | 3.7 | 8.5 | 8.1 |
| 154 | Pasadena | Texas | 2.1 | 4(80%) | 37783 | 3.9 | 6.9 | 7.8 |
| 6 | San Diego | California | 2.3 | 4(84%) | 30901 | 3.3 | 9.2 | 8.7 |
| 152 | Hollywood | Florida | 2.1 | 4(84%) | 22582 | 3.5 | 7.3 | 8.1 |
| 8 | Dallas | Texas | 2.0 | 5(84%) | 22491 | 4.3 | 6.9 | 6.9 |
| 61 | Arlington | Texas | 2.0 | 4(79%) | 21652 | 3.9 | 7.1 | 7.3 |
| 63 | Las Vegas | Nevada | 2.2 | 3(78%) | 19758 | 3.2 | 7.2 | 8.5 |
| 157 | Sunnyvale | California | 2.2 | 5(84%) | 18253 | 3.3 | 6.6 | 8.1 |
| 65 | St. Petersburg | Florida | 2.1 | 4(81%) | 16653 | 3.8 | 6.8 | 7.3 |
| 7 | Detroit | Michigan | 1.9 | 6(82%) | 16148 | 5.3 | 5.5 | 5.8 |
| 10 | San Antonio | Texas | 2.0 | 5(84%) | 16071 | 4.0 | 7.0 | 7.0 |
| 1 | New York | New York | 1.7 | 7(79%) | 15172 | 6.9 | 6.5 | 5.0 |
| 60 | Birmingham | Alabama | 2.2 | 4(82%) | 14987 | 3.5 | 8.9 | 7.8 |
| 95 | Tacoma | Washington | 2.4 | 3(80%) | 14221 | 3.1 | 9.3 | 8.3 |
| 156 | Sterling Heights | Michigan | 2.2 | 4(84%) | 10464 | 3.4 | 6.0 | 7.5 |
| 155 | Moreno Valley | California | 2.2 | 4(85%) | 9871 | 3.3 | 7.3 | 7.8 |
| 58 | Louisville | Kentucky | 2.2 | 4(83%) | 9862 | 3.5 | 6.7 | 7.4 |
| 89 | Dayton | Ohio | 2.1 | 4(80%) | 8113 | 3.7 | 6.9 | 6.8 |
| 5 | Philadelphia | Pennsylvania | 1.8/4.2 | 6(82%) | 7834 | 6.1 | 5.0 | 5.0 |
| 88 | Greensboro | North Carolina | 2.2 | 4(84%) | 6662 | 3.4 | 6.9 | 7.2 |
| 158 | Gary | Indiana | 2.0 | 5(84%) | 6365 | 4.1 | 7.5 | 6.3 |
| 96 | Little Rock | Arkansas | 2.1 | 4(79%) | 6160 | 3.9 | 8.8 | 6.4 |
| 94 | Spokane | Washington | 2.0 | 5(80%) | 5703 | 4.6 | 6.3 | 5.7 |
| 97 | Bakersfield | California | 2.1 | 4(82%) | 5582 | 3.6 | 6.3 | 6.7 |
| 56 | Newark | New Jersey | 2.0 | 5(83%) | 5485 | 4.4 | 5.9 | 5.8 |
| 57 | Saint Paul | Minnesota | 1.9 | 5(82%) | 5403 | 4.6 | 5.2 | 5.6 |
| 99 | Fort Wayne | Indiana | 2.2 | 4(82%) | 4668 | 3.5 | 6.3 | 6.7 |
| 98 | Fremont | California | 2.2 | 4(82%) | 4667 | 3.5 | 6.3 | 6.7 |
| 64 | Corpus Christi | Texas | 2.0 | 4(79%) | 3822 | 4.1 | 5.7 | 5.9 |
| 59 | Anaheim | California | 2.3 | 3(79%) | 3530 | 3.0 | 6.8 | 7.5 |
| 93 | Columbus | Georgia | 2.0 | 4(80%) | 2878 | 3.7 | 6.1 | 6.1 |
| 62 | Norfolk | Virginia | 1.9 | 5(79%) | 2808 | 4.5 | 5.3 | 5.3 |
| 153 | Topeka | Kansas | 2.0 | 5(83%) | 2656 | 4.2 | 5.2 | 5.5 |
| 159 | Beaumont | Texas | 1.9 | 5(80%) | 1911 | 4.7 | 5.3 | 4.9 |
| 100 | Arlington | Virginia | 2.0 | 5(84%) | 1703 | 4.2 | 4.7 | 5.2 |
| 151 | Laredo | Texas | 1.7/4.2 | 6(79%) | 1663 | 5.7 | 5.9 | 4.3 |
| 160 | Fullerton | California | 2.3 | 4(84%) | 1499 | 3.2 | 5.6 | 6.2 |

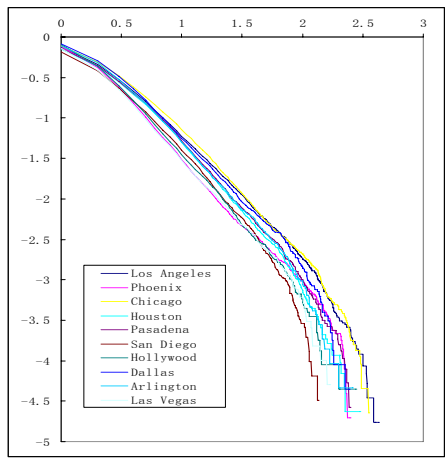
(a)

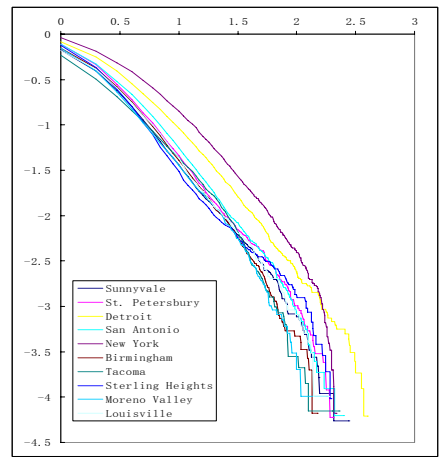
(b)

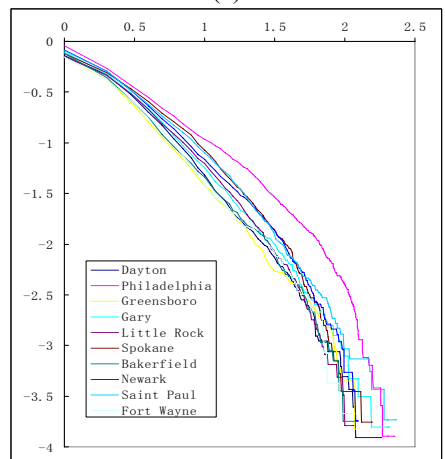
(c)

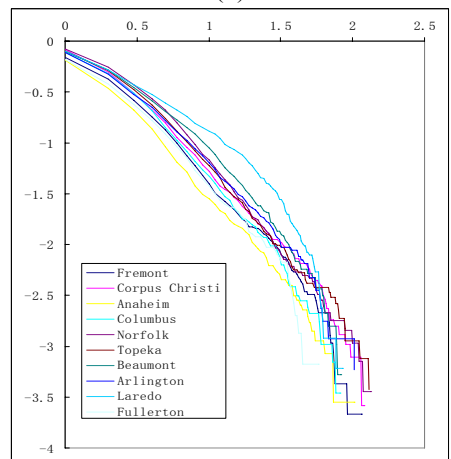
(d)

Figure 2: Log-log plots showing power law distributions with the four groups of cities with size ranges: (a) (19000 – 60000), (b) (9000 – 19000), (c) (4668 – 9000), and (d) (1499 – 4667)

The power law distributions indicate that most streets have very low degrees, while a few have extremely high degrees. As degree and path length are significantly correlated, i.e., lengthy streets tend to have more streets intersected with them, the power law is applicable to street length as well. This scaling property for both degree and street length can be illustrated in a detailed way shown in Figure 3. In words, this pattern states that about 80% of streets with a street network have length or degrees less than the average of the network, while 20% of streets have length or degrees greater than the average. Out of the 20%, there are less than 1% of streets which can form a backbone of the street network (Figure 4). In other words, all the streets are put around the average degree into two groups: those well connected (20%, with less than 1% extremely well connected), and those less connected (80%). Interestingly the pattern seems universal to all the cities studied.

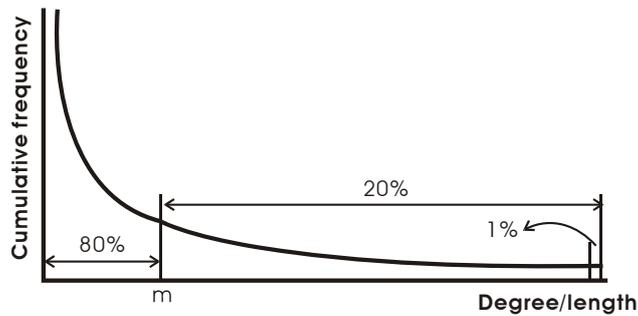

Figure 3: The topological pattern: 80% of streets having a degree less than the average m, while 20% having a degree greater than the average, out of the 20%, less than 1% of streets form a backbone of the street network.

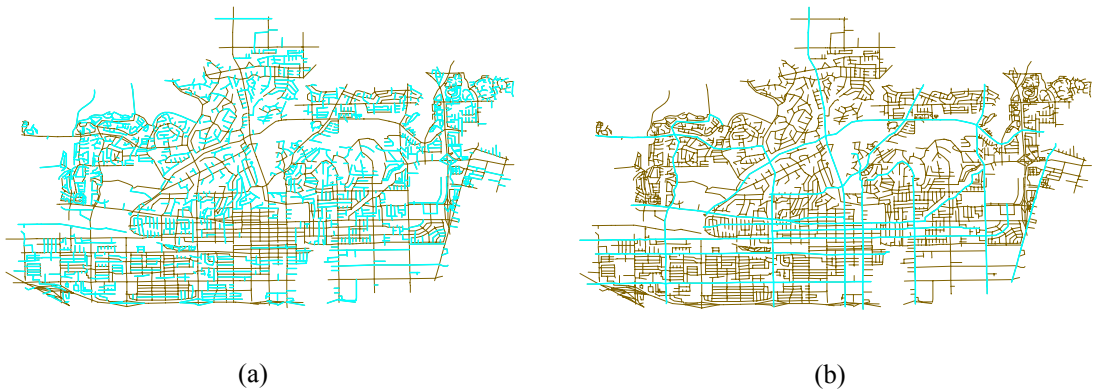

(a)                                                      (b)

Figure 4: 80% trivial streets highlighted with (a) and 1% vital streets highlighted with (b), which tend to form an image of the city (example of Fullerton)

The above topological pattern appears to be universal and this has been reconfirmed by 6 other cities (Table 2). Among the cities, Duffield in Virginia is the smallest U.S. town, according to the Guinness Book of Records [18]. To further verify the above pattern, we selected the parts of a city with different sizes (Figure 5) for the topological analysis. To our surprise, the selected parts of an urban street network still demonstrate the universal pattern. We should remark that when the number of streets selected is less than a couple of hundred, the pattern may disappear. This is not surprising, since the selected streets are too small to be representative, noting that the kind of topological analysis is for large networks with hundreds or thousands of vertices in essence.

Table 2: The other 6 urban street networks and their topological measures
(Rank = ranking according to population, α = power law curve exponent, partition = percentage around the average degree, n = number of streets, m = average degree of streets, $\ell$ = path length, $\ell\ rand$ = path length of the random counterpart)

| City | Country | α | Partition | n | m | $\ell$ | $\ell\ rand$ |
|---|---|---|---|---|---|---|---|
| Gävle | Sweden | 2.2 | 4(84%) | 1291 | 3.3 | 3.7 | 6 |

| Munich | Germany | 1.9 | 5(78%) | 831 | 4.6 | 4.7 | 4.4 |
| San Francisco | USA | 1.7/3.4 | 7(80%) | 2717 | 6.3 | 4.9 | 4.3 |
| Tel Aviv | Israel | 2.1 | 4(82%) | 3116 | 3.6 | 6.2 | 6.3 |
| Hong Kong | China | 2.1 | 4(82%) | 13429 | 3.5 | 9.7 | 7.5 |
| Duffield | USA | 2.6 | 3(83%) | 189 | 2.8 | 3.5 | 5.2 |

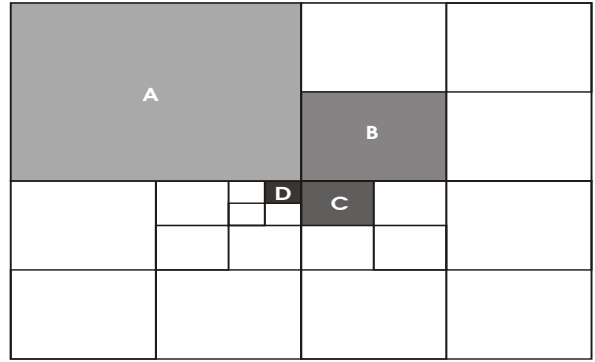

Figure 5: Selection rectangles (shaded)

Having illustrated the universality, now let's tend to the peculiarity in terms of the topological pattern. Taking the example of New York, Manhattan's grid-like structure is strikingly different from its neighboring areas. This can be seen clearly from the power law distributions of the different districts (Figure 6a), i.e., the log-log curve for Manhattan is not a straight line overall, and it deviates from the other curves. This is the peculiarity that we intend to illustrate. The reader may have also noticed that the power law exponent has two different values for cities including Philadelphia, Laredo and San Francisco in the above tables. We use the example of San Francisco to illustrate the fact. For San Francisco, it contains in total 2717 streets. Among them, 96.6% of streets have a degree less than 33, while the rest of the streets have a degree greater than 33. Those streets having a degree less than 33 follow the power distribution with exponent 1.7, while those streets having a degree greater than 33 follow the power distribution with exponent 3.4. This is clearly illustrated in Figure 6b.

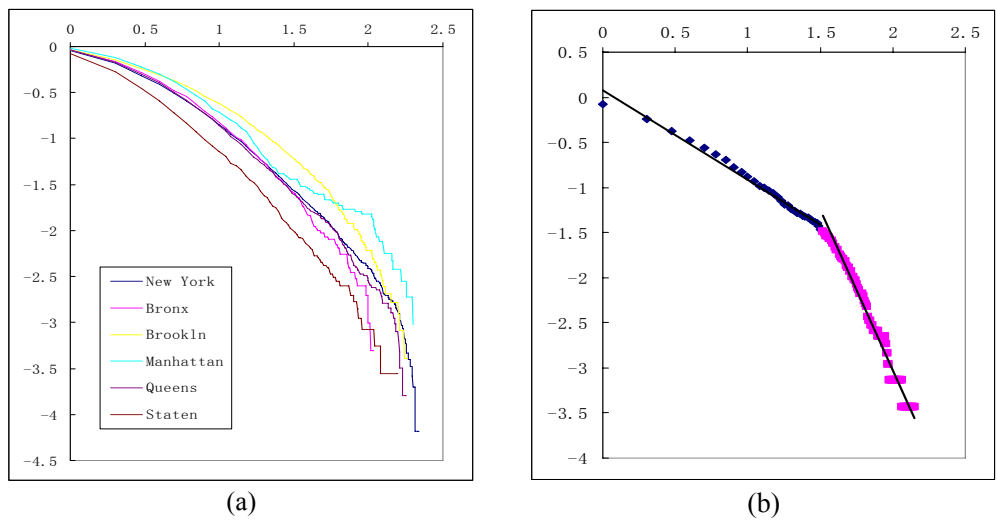

(a)    (b)

Figure 6: Peculiarity of power law distribution for New York and its districts (a), and double exponents for degree distribution of the San Francisco network (b)

This bipartite power law distribution implies that all the streets can be put into two parts, and each part follows a power law distribution with an exponent. The reason underlying the double exponents is still unclear, but we suspect it is due to grid-like street networks as both Manhattan and San Francisco have very similar grid-like patterns.

## 4. Conclusion

This paper studied street-street intersection topology using the largest data sample of urban street networks. We illustrated that the topologies of urban street networks are far from random, but small worlds with scale-free property. The scale-free property can be further described in detail as follows: about 80% of streets with a street network have length or degrees less than the average value of the network, while 20% of streets have length or degrees greater than the average. Out of the 20%, there are less than 1% of streets which can form a backbone of the street network. Thus, all the streets are put around the average degree into two unbalanced groups. Urban street networks on the surface have a pretty simple structure, but underneath or at the cognitive level is a consistent recurrence of this unbalanced pattern. This scaling pattern provides for the first time quantitative evidence as to why an image of the city [19] can be formed in our minds from the perspective of streets, i.e., the vital few tend to be landmarks in human cognitive maps. The pattern also helps to understand the robustness of urban street networks, i.e., the error tolerance to random "attacks" like congestion.

This inbuilt imbalance leads us to conjecture that 20% well connected streets account for 80% of traffic flow, and less than 1% extremely well connected streets constitute a cognitive map of an urban street network. If the conjecture does hold true, we do not need to invest equally in all streets, but the 20% to get a maximum reward – least effort in Zipf's term [20]. What "invisible hands" create urban street networks which all follow the power law distributions still seem unclear to us, but the illustrated pattern sheds substantial light on the notion of self-organizing cities [21, 22]. We tend to believe that the universal power law is a patent signature of self-organizing cities. In other words, cities are developed much in the same way as living organs like embryos. Apart from the findings illustrated, we have set up a comprehensive database of topologies of urban street networks which can be used as benchmark data for further research. To the best of our knowledge, it is the largest data sample for this kind of topological analysis, not only for urban street networks but also for other types of real world networks. The database will be put online for public access in the near future.

**Acknowledgement**
This work is supported by a Hong Kong Polytechnic University research grant. The author would like to thank Chengke Liu for his research assistance.